\documentclass[amsfonts,amssymb,twocolumn,showpacs]{revtex4}

\usepackage{graphicx}

\usepackage{amsmath}
\usepackage{amstext}
\usepackage{amssymb}

\begin{document}

\title{
BOLTZMANN-GIBBS ENTROPY IS SUFFICIENT BUT NOT NECESSARY FOR THE LIKELIHOOD FACTORIZATION REQUIRED BY EINSTEIN}

\author{
Constantino Tsallis$^{1,2}$ and Hans J. Haubold$^3$}
\affiliation{$^1$Centro Brasileiro de Pesquisas Fisicas and \\National Institute of Science and Technology for Complex Systems,
Rua Xavier Sigaud 150, 22290-180 Rio de Janeiro-RJ, Brazil\\
$^2$Santa Fe Institute, 1399 Hyde Park Road, Santa Fe, NM 87501, USA\\
$^3$Office for Outer Space Affairs, United Nations, P.O. Box 500, A-1400 Vienna, Austria}

\begin{abstract}
In 1910 Einstein published a crucial aspect of his understanding of Boltzmann entropy. He essentially argued that the likelihood function of any system composed by two probabilistically independent subsystems {\it ought} to be factorizable into the likelihood functions of each of the subsystems. Consistently he was satisfied by the fact that Boltzmann (additive) entropy fulfills this epistemologically fundamental requirement. We show here that entropies (e.g., the $q$-entropy on which nonextensive statistical mechanics is based) which generalize the BG one through violation of its well known additivity can {\it also} fulfill the same requirement. This fact sheds light on the very foundations of the connection between the micro- and macro-scopic worlds.
\end{abstract}


\maketitle

Einstein presented in 1910 \cite{Einstein1910} an interesting argument of why he liked Boltzmann's connection between the classical thermodynamic entropy introduced by Clausius and the probabilities of microscopic configurations. This argument is based on the {\it factorization} of the likelihood function of {\it independent} systems ($A$ and $B$), namely
\begin{equation}
\label{principle}
W(A+B)=W(A) W(B)
\end{equation}
This is a very powerful epistemological reason since it reflects the basic procedure of all sciences, namely that, in order to study any given natural, artificial and social system, we must start by ideally considering it isolated from the rest of the universe. In the present paper, we shall refer to Eq. (\ref{principle}), as {\it Einstein likelihood principle} (see \cite{Cohen20022005} for related aspects).

From Boltzmann celebrated principle 
\begin{equation}
S_{BG}=k \ln W \,,
\end{equation}
where $k$ is a conventional constant  (we shall from now on use $BG$, standing for {\it Boltzmann-Gibbs}, instead of just $B$), we obtain, through Einstein's well known {\it reversal} of Boltzmann formula, the likelihood function
\begin{equation}
\label{likelihood}
W \propto e^{S_{BG}/k} \,,
\end{equation}
with
\begin{equation}
S_{BG} = k \sum_i p_i \ln \frac{1}{p_i} \;\;\;\;\Bigl(\sum_{i=1}^W p_i =1\Bigr)              \,,
\end{equation}
where, for simplicity, we have used the case of discrete variables (instead of the continuous ones, that were of course used in the early times of statistical mechanics).
This entropy is {\it additive} according to Penrose's definition \cite{Penrose1970}. Indeed, if $A$ and $B$ are probabilistically {\it independent} systems (i.e., if $p_{ij}^{A+B}=p+i^A p_j^B$), we straightforwardly verify that
\begin{equation}
S_{BG}(A+B)=S_{BG}(A) + S_{BG}(B) \,,
\end{equation}
hence, by replacing this equality into Eq. (\ref{likelihood}), Eq. (\ref{principle}) is satisfied.

Let us exhibit now that entropic additivity is sufficient but not necessary for the Einstein principle (\ref{principle}) to be satisfied. Let us consider the following generalised functional \cite{Tsallis1988}, basis of nonextensive statistical mechanics \cite{Tsallis1988,GellMannTsallis2004,Tsallis2009,Tsallis2014},
\begin{eqnarray}
\label{qentropy}
S_q&=&k\frac{1-\sum_{i=1}^Wp_i^q}{q-1}  = k  \sum_{i=1}^W p_i \ln_q \frac{1}{p_i}       \\
&&\Bigl(q \in {\cal R}; \, \sum_{i=1}^W p_i=1;\,S_1=S_{BG} \Bigr), \nonumber 
\end{eqnarray}
with $\ln_q z \equiv \frac{z^{1-q}-1}{1-q}$  ($z>0$; $\ln_1 z=\ln z$). If $A$ and $B$ are two probabilistically independent systems (i.e., $p_{ij}^{A+B}=p_i^Ap_j^B$, $\forall (i,j)$), definition (\ref{qentropy}) implies 
\begin{eqnarray}
\frac{S_q(A+B)}{k} &=&  \frac{S_q(A)}{k}+ \frac{S_q(B)}{k} \nonumber \\
&&+ (1-q)\frac{S_q(A)}{k}\frac{S_q(B)}{k} \,.
\label{nonadditive}
\end{eqnarray} 
Consequently, according to the definition of entropic additivity in \cite{Penrose1970}, $S_q$ is additive if $q=1$, and {\it nonadditive} otherwise.

If probabilities are all equal, we straightforwardly obtain from (\ref{qentropy})
\begin{equation}
S_q=k \ln_q W \,,
\end{equation} 
hence Eq. (\ref{likelihood}) is generalised into
\begin{equation}
W \propto e_q^{S_{q}/k} \,,
\label{qlikelihood}
\end{equation}
where $e_q^z$ is the inverse function of $\ln_q z$ (hence, $e_q^z \equiv [1+(1-q)z]^{1/(1-q)}; \, e_1^z=e^z$).
If we take into account Eq. (\ref{nonadditive}), and use $e_q^{x\oplus_q y}=e_q^x e_q^y$ [with $x\oplus_q y \equiv x+y+(1-q)xy$, and $\ln_q (x y)=(\ln_q x) \oplus_q (\ln_q y)$], once again we easily verify Einstein's principle (\ref{principle}), but now {\it for arbitrary value of the index $q$} ! This exhibits a most important fact, namely {\it entropic additivity is not necessary for satisfying Einstein 1910 crucial requirement within the foundations of statistical mechanics.}   \\

In fact, this property is amazingly general. Indeed, let us consider a generalised trace-form entropic functional $S_G(\{p_i\}) \equiv  k  \sum_{i=1}^W p_i \ln_G \frac{1}{p_i}$, where $\ln_G z$ is some well behaved generalization of the standard logarithmic function. Let us further assume that, for probabilistically independent systems $A$ and $B$, $S_G$ satisfies $S_G(A+B)=\Phi (S_G(A),S_G(B)) \equiv S_G(A) \oplus_G S_G(B) $, where $\Phi$ denotes some generic function, and $\oplus_G$ generalises the standard sum. For equal probabilities (i.e., $p_i=1/W$), $S_G$ takes a specific form, namely $S_G(W) =k \ln_G W$. We shall note $e_G^z$ the inverse function of $\ln_G z$. 
Then, following Einstein's reversal, the likelihood function is given by  
\begin{equation}
W \propto e_G^{S_{G}/k} \,,
\label{Glikelihood}
\end{equation}
and, once again, by using $e_G^{x \oplus_G y}=e_G^{x}e_G^{y}$, the Einstein principle (\ref{principle}) is satisfied, $\forall G$. Clearly, the additive $S_{BG}$ and the nonadditive $S_q$ are particular illustrations of this property. Another example which follows this path is the equal-probability case of another, recently introduced (to address black holes    \cite{TsallisCirto2013,Bekenstein1973,Hawking1974,tHooft1990,Susskind1993,Maddox1993,Srednicki1993,StromingerVafa1996,Strominger1998,MaldacenaStrominger1998,Padmanabhan2009,Corda2011} and the so-called area law \cite{EisertCramerPlenio2010}), 
nonadditive entropy, namely (footnote of page 69 in \cite{Tsallis2009}, and \cite{TsallisCirto2013}; see also \cite{Ubriaco2009}),  
\begin{equation}
\label{deltaentropy}
S_\delta=k_B \sum_{i=1}^W p_i \Bigl(\ln\frac{1}{p_i} \Bigr)^\delta \;\;\; (\delta > 0; \, S_1=S_{BG}) \,.
\end{equation}
For equal probabilities we have
\begin{equation}
\label{deltaentropy2} 
S_\delta = k_B \ln^\delta W \,,
\end{equation}
hence, for $\delta >0$,
\begin{eqnarray}
\frac{S_\delta(A+B)}{k_B} &=&  \Bigl\{   \Bigl[ \frac{S_\delta(A)}{k_B} \Bigr]^{1/\delta} +     \Bigl[  \frac{S_\delta(B)}{k_B} \Bigr]^{1/\delta}   \Bigr\}^\delta \nonumber \\
&\equiv& \frac{S_\delta(A)}{k} \oplus_\delta  \frac{S_\delta(B)}{k}
\end{eqnarray}

Let us note at this point a crucial issue, namely that entropies $S_{BG}$, $S_q$ and $S_\delta$ are thermodynamically appropriate for systems constituted by $N$ elements, such that the total number of admissible microscopic configurations are, in the $N \to\infty$ limit, given respectively by 
$C \mu^N \;\;\; (C>0;\, \mu>1)$, $D N^\rho \;\;\; (D>0; \, \rho>0)$ and $\phi(N)\nu^{N^\gamma} \;\;(\nu>1; \, 0<\gamma<1)$ [$\phi(N)$ being any function satisfying $\lim_{N\to\infty} \frac{\ln \phi(N)}{N^\gamma}=0$; strictly speaking, $C$ and $D$ could also be sufficiently slowly varying functions of $N$].  Notice that $C \mu^N >> \phi(N)\nu^{N^\gamma} >> D N^\rho$, which implies that the Lebesgue measure of the phase-space occupancy typically {\it vanishes} for the cases where nonadditive entropies are to be used, whereas it is {\it nonzero} in the standard BG case. In all cases, for special values of $q$ (namely $q=1-1/\rho$) or $\delta$ (namely $\delta = 1/\gamma$), the thermodynamical requirement that $S(N) \propto N$ is satisfied!
It is possible to unify the entire discussion by defining \cite{TsallisCirto2013} $S_{q,\delta}
= k_B  \sum_{i=1}^W p_i \Bigl(\ln_q \frac{1}{p_i}\Bigr)^\delta     \;    \Bigl(q \in {\cal R}; \, \delta>0 \Bigr)$. Indeed, $S_{1,1}=S_{BG}$, $S_{q,1}=S_{q}$, and  $S_{1,\delta}=S_{\delta}$.

\begin{figure}
\begin{center}
\includegraphics[width=8.2cm]{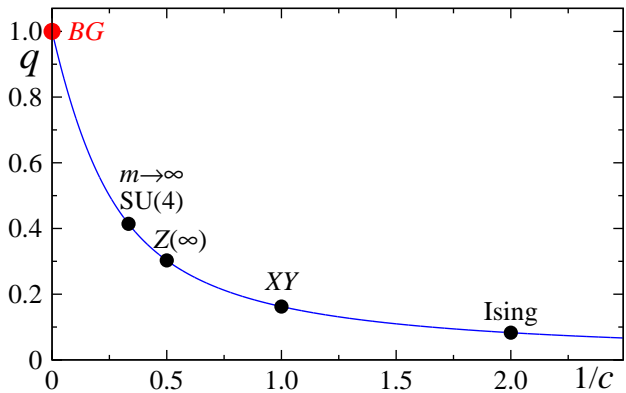}
\end{center}
\caption{The index $q$ has been determined \cite{CarusoTsallis2008} from first principles, namely from the universality class of the Hamiltonian. The values $c=1/2$ and $c=1$ respectively correspond to the Ising and XY ferromagnetic chains in the presence of transverse field at $T=0$ criticality. For other models see \cite{Alcaraz}.
In the $c \to\infty$ limit we recover the Boltzmann-Gibbs (BG) value, i.e.,  $q=1$. For arbitrary value of $c$, the subsystem {\it nonadditive} entropy $S_q$ is thermodynamically {\it extensive} for, and only for,  $q=\frac{\sqrt{9+c^2}-3}{c}$.}   
\label{Caruso}
\end{figure}

\begin{figure}
\begin{center}
\includegraphics[width=8.2cm]
{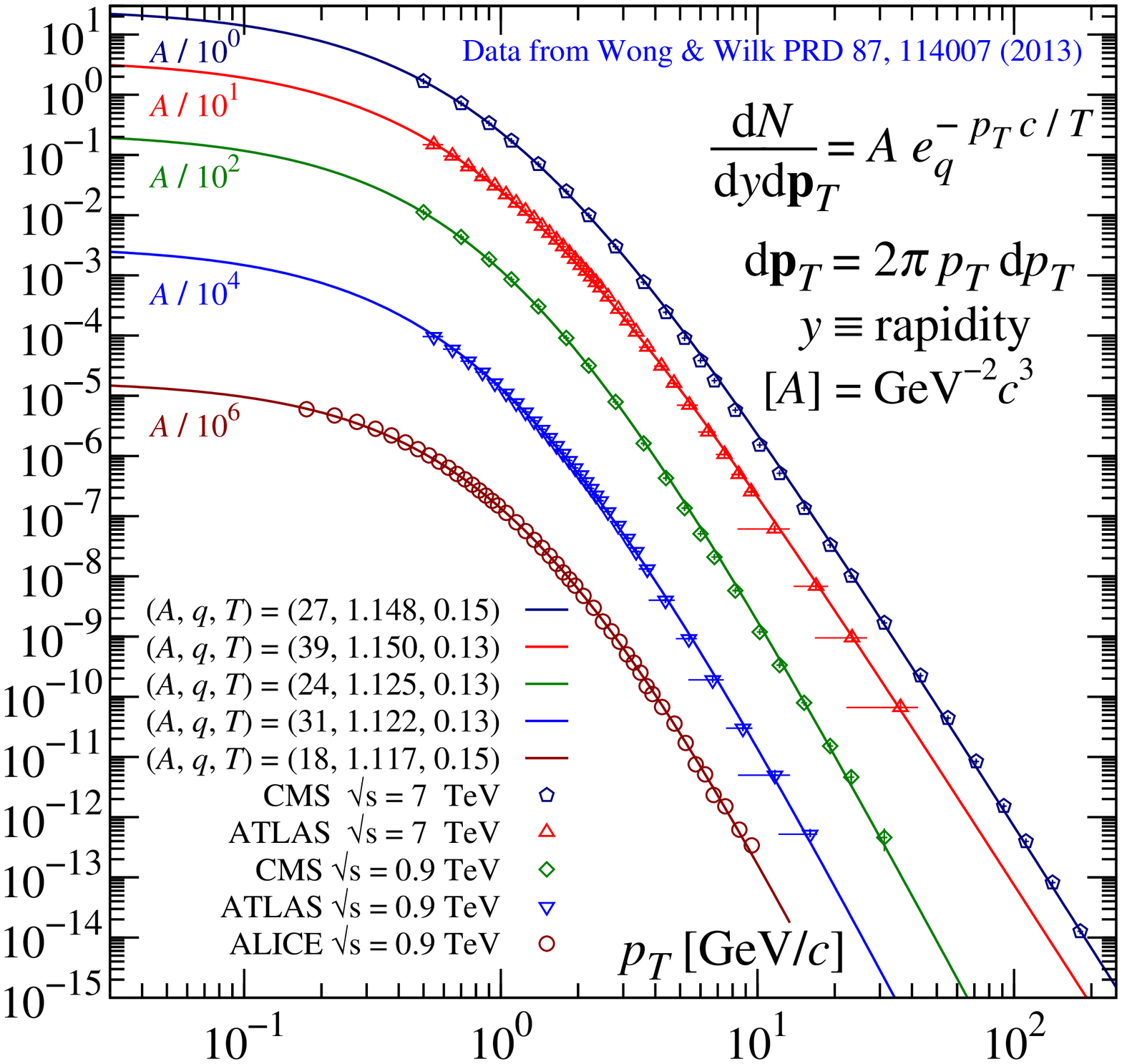}
\end{center}
\caption{Experimental distributions of the transverse moments in hadronic jets at the CMS, ALICE and ATLAS detectors at LHC. The data are from \cite{WongWilk2013}. They can be remarkably well fitted (along fourteen decades) with the $q$-exponential function $e_q^x \equiv [1+(1-q)x]^{1/(1-q)}$, which, under appropriate constraints, extremises the entropy $S_q$. See details in \cite{CirtoTsallisWongWilk2014}.}   
\label{LHC}
\end{figure}

It is fundamental to stress that indices such as $q$ and $\delta$ are to be to obtained from first principles, i.e., from mechanics (classical, quantum, relativistic). This is already seen by the fact that, in the two above illustrations, $q$ (or $\delta$) is obtained directly from $\rho$ (or $\gamma$). This means that, if we are dealing say with Hamiltonian systems, $q$ and $\delta$ are in principle determined directly from the Hamiltonian, more precisely from the universality class of the Hamiltonian. One paradigmatic nontrivial illustration is analytically available in the literature \cite{CarusoTsallis2008}. It concerns the entropy of a thermodynamically large subsystem of a strongly quantum entangled one-dimensional many-body system which belongs to the universality class characterized by the {\it central charge} $c$. Indeed, at quantum criticality (i.e., at $T=0$), we have $q=\frac{\sqrt{9+c^2}-3}{c}$: see Fig. \ref{Caruso}. It is clear that, in contrast with this example, the analytical determination of $q$ appears to be mathematically intractable for most systems. This is the only reason why we plethorically find in the literature papers where the indices $q$ are determined through fitting procedures. In some examples, the fitting can nevertheless be amazingly precise: see \cite{WongWilk2013,CirtoTsallisWongWilk2014} and Fig. \ref{LHC}.

Complexity frequently emerges in natural, artificial and social systems. It may be caused by various geometrical-dynamical ingredients, which include non-ergodicity, long-term memory, multifractality, and other spatial-temporal long-range correlations between the elements of the system, which ultimately drastically restrict the total number of microscopically  admissible possibilities. During the last two decades, many such phenomena have been successfully approached in the frame of nonadditive entropies and nonextensive statistical mechanics. Predictions, verifications and various applications have been performed in high-energy physics \cite{CMS1},
spin-glasses \cite{PickupCywinskiPappasFaragoFouquet2009}, cold atoms in optical lattices \cite{DouglasBergaminiRenzoni2006}, trapped ions \cite{DeVoe2009},
anomalous diffusion \cite{AndradeSilvaMoreiraNobreCurado2010}, dusty plasmas \cite{LiuGoree2008},  solar physics \cite{BurlagaVinasNessAcuna2006}, long-range interactions \cite{CirtoAssisTsallis2014},  
relativistic and nonrelativistic nonlinear quantum mechanics \cite{NobreMonteiroTsallis2011}, among many others.

All this is totally consistent with the fact that, for all those systems for which the correlations between the microscopic degrees of freedom is generically weak, the thermodynamically admissible entropy is precisely the additive one, $S_{BG}$, as well known. If, however, strong correlations are generically present (e.g., of the type assumed in the $q$-generalization of the Central Limit and L\'evy-Gnedenko Theorems \cite{UmarovTsallisSteinberg2008}), we need to implement nonadditive entropies \cite{TsallisGellMannSato2005,HanelThurner2011,RuizTsallis2011} and their associated statistical mechanics.

Summarizing, in order to satisfy classical thermodynamics, the thermostatistics of a wide class of systems whose elements are strongly correlated (for instance, through long-range interactions, or through strong quantum entanglement, or both, like possibly in quantum gravitational dense systems) are to be based on nonadditive entropies such as $S_{q, \delta}$ \cite{REMARK}, and not on the usual Boltzmann-Gibbs-von Neumann one. Nevertheless, and this is the main point of the present note, {\it Einstein's principle} (\ref{principle}) {\it is generically satisfied for a wide class of entropies (which includes $S_q$ and others) and not only for the BG one}.\\

\section*{Acknowledgments}
\vspace{-0.4cm}
We have benefitted from interesting remarks by L.J.L. Cirto, A.M. Mathai, G. Wilk and C.Y. Wong. Partial financial support from CNPq, Faperj and Capes (Brazilian agencies) is acknowledged as well.

\end{document}